\documentclass[useAMS,usenatbib,letters]{mn2e}

\include{epsf}
\newcommand{\be}{\begin{equation}}
\newcommand{\ee}{\end{equation}}
\newcommand{\ba}{\begin{eqnarray}}
\newcommand{\ea}{\end{eqnarray}}

\def\simless{\mathbin{\lower 3pt\hbox
   {$\rlap{\raise 4pt\hbox{$\char'074$}}\mathchar"7218$}}}
\def\simgreat{\mathbin{\lower 3pt\hbox
   {$\rlap{\raise 4pt\hbox{$\char'076$}}\mathchar"7218$}}}
\title[SFH evolution of BLAST galaxies]
{Evolution of the star formation histories of BLAST galaxies} 
\author[Simon Dye]{ 
\parbox[t]{\textwidth}{
Simon Dye$^1$\thanks{E-mail: s.dye@astro.cf.ac.uk},
Steve Eales$^1$,
Lorenzo Moncelsi$^1$,
Enzo Pascale$^1$
}\\ \\
$^1$Cardiff University, School of Physics \& Astronomy, Queens Buildings,
The Parade, Cardiff, CF24 3AA, U.K.
}

\begin{document}

\date{Document in prep.}

\pagerange{\pageref{firstpage}--\pageref{lastpage}} \pubyear{2010}

\maketitle

\label{firstpage}

\begin{abstract}

We have measured star formation histories (SFHs) and stellar masses of
galaxies detected by the Balloon-borne Large Aperture Sub-millimetre
Telescope (BLAST) over $\sim 9$\,deg$^2$ centred on the Chandra Deep
Field South. We have applied the recently developed SFH reconstruction
method of Dye et al. to optical, near-infrared and mid-infrared
photometry of 92 BLAST galaxies. We find significant differences
between the SFHs of low mass ($\simless 10^{11}$\,M$_\odot$) and high
mass ($\simgreat 10^{11}$\,M$_\odot$) systems. On average, low mass
systems exhibit a dominant late burst of star formation which creates
a large fraction of their stellar mass. Conversely, high mass systems
tend to have a significant amount of stellar mass that formed much
earlier. We also find that the high mass SFHs evolve more strongly
than the low mass SFHs. These findings are consistent with the
phenomenon of downsizing observed in optically selected samples of
galaxies.

\end{abstract}


\section{Introduction}
\label{sec_intro}

Encoded in every galaxy's spectrum is a record of its entire life from
birth, up to the epoch at which it is observed. Unlocking this
information, by determining the variation in star formation rate (SFR)
with age, is a key step towards understanding how galaxies form and
evolve. The measurement of star formation histories (SFHs) therefore
plays a crucial role in the development of an accurate model to
describe the range of processes experienced by galaxies and the
subsequent formation of stellar mass in the Universe.

Currently, the most elusive population of galaxies are systems
heavily obscured by dust, often detected only at sub-millimetre
(submm) wavelengths. Approximately half of all light emitted by
galaxies is absorbed by dust and re-radiated in the submm. However,
compared with studies at optical wavelengths, little is known about
submm galaxies, in particular, how they relate to local systems. Submm
selected samples of galaxies therefore provide an unavoidably
important set of constraints on a complete and self-consistent view of
galaxy formation mechanisms.

\citep{dye_et_al08} conducted an analysis of the SFHs of 850$\,\mu$m
selected galaxies and found that these systems are typically dominated
by a strong burst of star formation late in their history, but that
around half of their stellar population had already formed over the
first half of their lives. This study also revealed a surprising
deficit of high mass ($> 5\times 10^{11} {\rm M}_\odot$) systems at
redshifts $z<2$, strong evidence of downsizing in the population,
possibly explained by the evolution of these systems into massive
ellipticals.

The Balloon-borne Large Aperture Sub-millimetre Telescope (BLAST)
recently provided a catalogue of hundreds of submm selected galaxies
over $\sim 9\,$deg$^2$ of sky \citep{devlin09} centred on the Chandra
Deep Field South (CDFS).  The identification of radio and 24$\,\mu$m
counterparts to the majority of sources \citep{dye09} combined with
follow-up optical spectroscopy \citep{eales09} and a comprehensive
suite of archival optical, near-infrared (near-IR) and mid-infrared
(mid-IR) imaging results in these data being the largest, most
thoroughly characterised and carefully processed sample of
250-500$\,\mu$m selected sources to date. The sample and its
supporting multi-wavelength data therefore presents a perfect
opportunity to conduct a study of SFHs of submm selected galaxies.

The purpose of this letter is therefore to carry out an investigation
of the SFHs and stellar masses of BLAST galaxies in a similar vein to
the study of 850$\,\mu$m selected galaxies conducted by
\citet{dye_et_al08}. To compute SFHs, we have used the recently
developed method of \citet{dye08}. We have applied the method to
optical, near-IR and mid-IR photometry of the counterparts to the
BLAST sources identified by \citet{dye09}.

In Section \ref{sec_method} we briefly outline the procedure used for
applying the SFH reconstruction method. Section \ref{sec_data}
describes the data. The results are presented in Section
\ref{sec_results}. Finally, we summarise in Section \ref{sec_summary}.
Throughout this letter, the following cosmological parameters are
assumed; ${\rm H}_0=71\,{\rm km\,s}^{-1}\,{\rm
Mpc}^{-1}$, $\Omega_m=0.27$, $\Omega_{\Lambda}=0.73$.

\section{SFH Reconstruction Method}
\label{sec_method}

\citet{dye08} provides a detailed description of the SFH
reconstruction method. The purpose of the brief outline presented
here is both for completeness and to give specific details of
the procedure we have used in this implementation.

\subsection{Method Outline}
\label{sec_outline}

The method divides a galaxy's history into discrete blocks of time.
This results in a relatively low resolution SFH, but one that does not
adhere to a prescribed (i.e., potentially biased) parametric
form. Using a synthetic library of simple stellar population (SSP)
SEDs, the fluxes resulting from a constant SFR normalised to one solar
mass in each block, as measured in the observer frame across a range
of filters, are calculated. Finding the contribution of flux from each
block in each filter that best fits a set of observed fluxes is a
linear problem that can be solved exactly. The solution directly
yields the galaxy's SFH and stellar mass.  In this letter, we have
used the stellar SED libraries of \citet{maraston05} and, for
comparison, \citet{bruzual03}.

Starting with a SSP SED, $L_{\lambda}^{\rm SSP}$, of metallicity $Z$,
a composite stellar population (CSP) SED, $L_{\lambda}^{\, i}$, is
generated for the $i$th block of constant star formation in a given
galaxy using
\be
L_{\lambda}^{\, i} = \frac{1}{\Delta t_i} \int^{t_i}_{t_{i-1}} {\rm d}t' \, 
L_{\lambda}^{\rm SSP}(\tau(z)-t')
\ee
where the block spans the period $t_{i-1}$ to $t_i$ in the galaxy's
history and $\tau$ is the age of the galaxy (i.e., the age of the
Universe today minus the look-back time to the galaxy).  The quantity
$\Delta t_i=t_i-t_{i-1}$ ensures that the CSP is normalised to one
solar mass.  The above integral is evaluated by interpolating linearly
in log($t$) between the discrete time intervals at which the SSP SEDs
are given in the libraries. Note that the method assumes that $Z$ does
not vary with age.

To model the effects of extinction on the final SED (i.e., the SED
from all blocks in the SFH), reddening is applied.  This is achieved
by individually reddening the CSP of each block using
$L_{\lambda,R}^{\, i} = L_{\lambda}^{\, i} \, 10^{-0.4
k(\lambda)A_V/R_V}$ where, $A_V$ is the extinction, $R_V=4.05$ and the
Calzetti Law \citep{calzetti00} is used for $k(\lambda)$.  The model
flux (i.e., photon count) observed in filter $j$ from a given block
$i$ in the SFH when the galaxy lies at a redshift $z$ is then
\be
F_{ij}=\frac{1}{4\pi d_L^{\,2}}\int {\rm d}\lambda 
\frac{\lambda \, L^i_{\lambda,R}(\lambda/(1+z))T_j(\lambda)}{(1+z)\, hc} 
\ee
where $d_L$ is the luminosity distance and $T_j$ is the transmission
curve of filter $j$. 

To find the normalisations $a_i$ which result in a set of model fluxes
that best fits the observed fluxes, the following $\chi^2$ function
is minimised
\be
\label{eq_chi_sq}
\chi^2=\sum_j^{N_{\rm filt}} \frac{(\sum_i^{N_{\rm block}} \,
a_i F_{ij} - F^{\rm obs}_j)^2}{\sigma_j^2}
\ee
where $F^{\rm obs}_j$ is the galaxy flux observed in filter $j$ and
$\sigma_j$ is its error. The sum in $i$ acts over all $N_{\rm block}$
SFH blocks. The minimum $\chi^2$ occurs when the condition
$\partial\chi^2/\partial\,a_i=0$ is simultaneously satisfied for all
$a_i$. This gives a set of equations linear in the $a_i$ which are
solvable using a standard matrix inversion \citep[see][for more
details]{dye08}. The $a_i$ are the stellar masses formed in each block
so that the total stellar mass of the galaxy is the sum ${\rm M}_* =
\sum_i^{N_{\rm block}} a_i$. The SFRs and hence the SFH is then given
directly by dividing the $a_i$ by the time spanned by each
corresponding block. 
Formal errors on the $a_i$ are obtained from the covariance
matrix, computed in a simple additional step. To allow for uncertainty
in source redshift, we performed a Monte Carlo simulation, randomising
the redshift according to its error, and combined the resulting
scatter in the $a_i$ in quadrature with the formal errors. The total
error on the stellar mass was obtained in the same manner.

As discussed in \citet{dye08}, regularisation must be
applied to ensure that the linear solution is well defined. The
strength of regularisation is controlled by a parameter
referred to as the regularisation weight, denoted $w$ hereafter.

\subsection{Fitting Procedure}
\label{sec_fitting}

The procedure outlined in the previous section is a single linear step
which computes the SFH that gives the best fit (i.e., minimum
$\chi^2$) to an observed set of fluxes for a given set of parameters
$z$, $Z$, $A_V$, $N_{\rm block}$ and $w$. This step is nested inside a
non-linear search for the set of parameter values that gives the best
overall fit to the observed fluxes. As discussed in \citet{dye08},
finding the best global solution can not be achieved by minimising
$\chi^2$ because the effective number of degrees of freedom depends on
$w$ in an unquantifiable manner. It is therefore not possible to use
$\chi^2$ to make a fair comparison of the goodness-of-fit between two
parameter sets with differing values of $w$. To make a fair
comparison, one must turn to Bayesian statistics and treat
regularisation as a prior. In this way, the Bayesian evidence, denoted
$\epsilon$ hereafter, allows different sets of parameters to be ranked
fairly. The best global solution is that which maximises $\epsilon$.

We investigated a range of different schemes to maximise $\epsilon$.
The most efficient and reliable scheme that we found combines a
standard grid search with a downhill simplex minimisation to find the
minimum of $-\ln \epsilon$ (hence the maximum of $\epsilon$).  We use
the simplex routine, linearly computing the SFH and evaluating
$\epsilon$ each time, to find the best pair of values of $A_V$ and
$\log w$ whilst keeping $Z$ and $N_{\rm block}$ fixed.  Then, at the
outer-most level, we step through a grid of regularly spaced trial
values of $\log Z$ and $N_{\rm block}$ over the ranges $ -2 <
\log_{10} (Z/Z_\odot) < 0.3 $ and $2 < N_{\rm block} < 5$. The value
of $z$ was fixed at the measured redshift of the galaxy (see
Section \ref{sec_data}) at all times.

\section{Data}
\label{sec_data}

\citet{dye09} found optical counterparts to 114 of the $\sim 130$
radio- and/or 24$\,\mu$m-identified BLAST sources located in the
region covered by their optical data. All 114 sources were detected at
$\geq 5\sigma$ in at least one of the three BLAST bands (250, 350 and
500$\,\mu$m). As described below, we acquired a range of optical,
near-IR and mid-IR photometry for these 114 sources. We rejected 12
sources on the basis of having less than five photometric data points,
applying our analysis to the remaining 102. In all cases, we used
total magnitudes/fluxes.

Optical photometry was taken from either the Spitzer Wide-area
InfraRed Extragalactic survey \citep[SWIRE;][]{lonsdale04}, or the 17
band COMBO-17 (Classifying Objects by Medium-Band Observations in 17
filters) survey \citep{wolf04}. The SWIRE survey covers $\sim
5$\,deg$^2$ to a depth of $r\simeq24.5$ (Vega, $5 \sigma$) and
COMBO-17 covers $\sim 0.25$\,deg$^2$ to $R\simeq26.0$ (Vega, $5
\sigma$).  The SWIRE optical catalogue directly provides total
magnitudes. For COMBO-17, we computed total magnitudes from aperture
fluxes in all filters using the given source-specific correction
derived in the $R$ band. For objects detected in both surveys, we
amalgamated both sets of photometry.

For the near-IR, we used the $J$ and $K$ photometry from the
Multi-wavelength Survey by Yale-Chile \citep[MUSYC][]{gawiser06}.
This reaches depths of 22.1 and 20.5 (Vega, 5$\sigma$) in $J$ and
$K$ respectively.  Although MUSYC only covers $\sim 0.3$\,deg$^2$, the
survey area is centred on the CDFS where the deep BLAST observations
were made \citep[see][]{dye09}, hence we obtained near-IR photometry
for 52 of the 102 sources. The MUSYC catalogue gives aperture and
total $K$ band fluxes. We obtained total $J$ band fluxes by scaling
the aperture fluxes by the total:aperture $K$ band flux ratio.

Mid-IR photometry was taken from the SWIRE survey which
completely encompasses the region containing the BLAST sources
considered in the present work. We used 3.6 and 4.5$\,\mu$m fluxes
which are limited to a 10$\sigma$ sensitivity of 10$\,\mu$Jy at
3.6$\,\mu$m and a 5$\sigma$ sensitivity of 10$\,\mu$Jy at 4.5$\,\mu$m.
For reconstructing source SFHs, we only used 3.6$\,\mu$m photometry
for objects with redshifts $z>0.6$ and additionally 4.5$\,\mu$m
photometry only for objects at $z>1$. These redshift limits ensure
that the observed photometry does not extend beyond the restframe $K$
band where the empirical stellar component of the Maraston SEDs ends.

We imposed a minimum photometric error of 0.05\,mag for all photometry
to allow for zero point uncertainties and mismatches in calibration
between different datasets.

In terms of redshifts, 47 of the sources have spectroscopic redshifts
taken from \citet{moncelsi10}. For the remainder, we used the photometric
redshifts assigned in \citet{dye09}. These are taken from either
\citet{mrr08} which uses the SWIRE optical photometry as well as the
SWIRE 3.6 and 4.5$\,\mu$m data or from COMBO-17.

\section{Results}
\label{sec_results}

We applied the procedure outlined in Section \ref{sec_fitting} to
maximise the Bayesian evidence for each galaxy. We found that 5-10\%
of reconstructions resulted in very poor fits to the observed
photometry and/or gave unphysical SFHs (i.e., strongly negative bursts
of star formation). These failures were readily discarded by applying a
cut in the evidence of $\ln \, \epsilon > -100$. This removed 10 of
the 102 sources, including three low redshift galaxies ($z<0.1$) which
are optically well resolved (up to $\sim 50''$ in diameter) and
therefore susceptible to strong photometric biases between the
different catalogues.  All analysis that follows in this letter is
applied to the remaining 92 systems.

\begin{figure}
\epsfxsize=83mm
{\hfill
\epsfbox{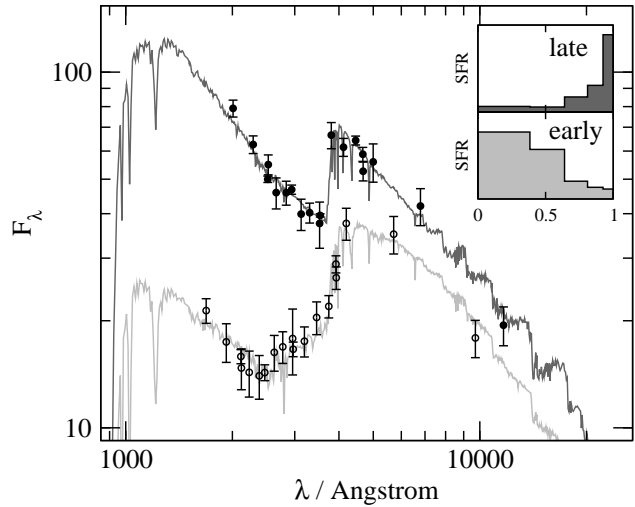}
\hfill}
\caption{Example rest-frame SEDs fitted to optical and near-IR
photometry of two BLAST sources. The darker SED, corresponding to a
$z\simeq0.8$ galaxy with observed photometry plotted as filled
photometric data points, is fit with an SFH where 60\% of its stellar
mass was formed in the last 10\% of its history and 15\% formed in the
first half.  The lighter SED, corresponding to a $z\simeq 1.2$ galaxy
with observed photometry plotted as unfilled data points, has a SFH
with only 3\% of its stellar mass formed in the last 10\% of its
history and 85\% formed in the first half.  Fluxes are in arbitrary
units and SEDs are independently scaled for clarity. Inset panels show
the SFR versus fractional galaxy age from birth ($t=0$) to the epoch
its observed ($t=1$) for each SED.
}
\label{SEDs}
\end{figure}

Our findings indicate that the vast majority of BLAST galaxies have
SFHs which peak at late times. Around 10\% of systems are more
consistent with early type SEDs having formed at least three quarters
of their stellar mass in the first half of their lives and less than
2\% of their mass in the last tenth of their lives. Figure \ref{SEDs}
shows two fairly extreme example SEDs of BLAST sources, one dominated
by late star formation and the other having formed nearly all of its
stars early on, to illustrate the range SEDs observed. 

\begin{figure}
\epsfxsize=83mm
{\hfill
\epsfbox{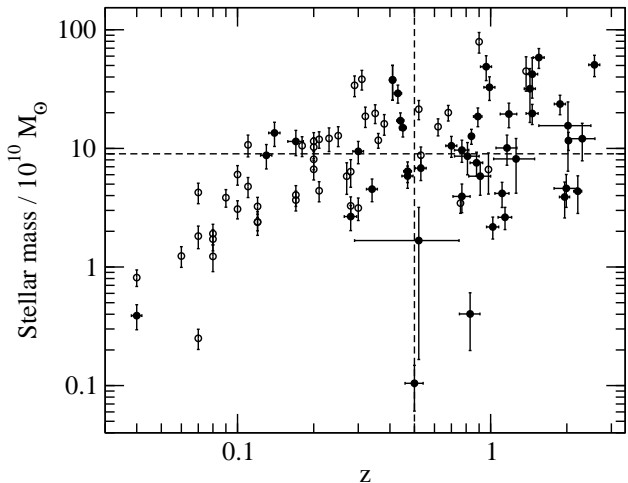}
\hfill}
\caption{Stellar mass in units $10^{10}\,$M$_\odot$ versus redshift
for the BLAST sources. The dashed lines at $z=0.5$ and
$9\times10^{10}\,$M$_\odot$ divide the sample into four approximately
equal sub-samples which we have then used in our analysis of 
SFH evolution by mass. Open points indicate sources where a
spectroscopic redshift has been used and filled points with errors
show the remaining sources with photometric redshifts. Mass errors
include photometric and redshift uncertainties.}
\label{mass_vs_z}
\end{figure}

Figure \ref{mass_vs_z} shows how the stellar masses vary with
redshift. There is an obvious trend of increasing mass with increasing
redshift, limited by the rarity of high mass galaxies at one end of
the scale and the sensitivity to low luminosity at the other. The few
low mass outliers are systems dominated by extremely late and intense
bursts of star formation and hence have low mass-to-light ratios.
Four of the 92 objects are flagged as having a dominant active
galactic nuclear component by \citet{moncelsi10} based on their
spectral line strength ratio [NII]658.3/H$\alpha$. These four objects
are all located near the high mass edge of the envelope shown in
Figure \ref{mass_vs_z} and spread throughout the redshift range.

\begin{figure}
\epsfxsize=83mm
{\hfill
\epsfbox{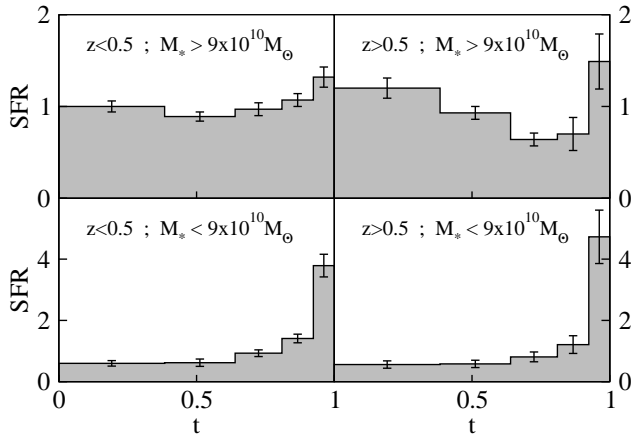}
\hfill}
\caption{Mean SFHs of the four sub-samples of BLAST galaxies as
delineated by the redshift and stellar mass limits indicated (see
Figure \ref{mass_vs_z}). 
The error bars on each histogram bin indicate
the $1\sigma$ error on the mean which includes the formal error
from the linear inversion and the uncertainty in source redshift computed
in the Monte Carlo analysis (see Section \ref{sec_outline}).
All SFRs are expressed in units of
fraction of total stellar mass per fractional age (i.e., the area
under each histogram in these units is 1). The fractional age varies
from $t=0$ at $z=\infty$ to $t=1$ at the epoch of the galaxy's
redshift.}
\label{SFHs}
\end{figure}

We investigated the possibility of evolution in the SFHs and whether
this depends on mass by segregating the sample of 92 galaxies into
four approximately equally sized sub-samples divided at $z=0.5$ and
$M_*=9\times10^{10}$\,M$_\odot$.  Figure \ref{SFHs} shows the average
SFH for each sub-sample, rebinned to a common resolution of five
blocks. In the averaging, we normalised the SFH of each galaxy to
units of fractional total stellar mass formed per fractional age so
that the integral of SFH over fractional age (i.e., the total stellar
mass) is equal to unity.

The most striking feature seen in the plots is the difference between
the low and high mass sub-samples. On average, both low mass SFHs are
dominated by a late burst of strong star formation activity accounting
for $\sim 40\%$ of the total stellar mass. Conversely, both high mass
SFHs show that a much smaller fraction of stellar mass ($\sim 5 -
10\%$) is created during this last period, the majority of mass being
formed at earlier times.

Another obvious effect seen in Figure \ref{SFHs} is that the high mass
sources exhibit a more prominent difference in their SFHs in moving
from high to low redshifts than the low mass sources. The high mass
sources have therefore, by this definition, undergone more evolution.
To quantify the significance of this, we computed the reduced $\chi^2$
statistic between the low and high redshift SFHs for the low and high
mass sub-samples in turn. For the high mass SFHs, the statistic is
$\chi^2_r=3.57\pm0.63$ compared to $\chi^2_r=0.51\pm0.63$ for the low
mass SFHs. The change at high mass is therefore significant at the
$\sim 3 \sigma$ level whereas the low mass source SFHs are consistent
with no change. This is synonymous with downsizing where the
instantaneous star formation rate in high mass galaxies evolves more
strongly than that in low mass systems \citep[e.g.,][]{heavens04}.

The SFHs computed in terms of fractional mass and fractional galaxy
age are a very useful diagnostic since they effectively normalise out
the large scatter in mass and redshift present in the necessarily
coarsely binned sub-samples. This makes the mean trends more
conspicuous.  However, to compare with more traditional studies of the
evolution of star formation, we estimated instantaneous absolute
SFRs. For each source, we computed a `pseudo-instantaneous' SFR by
dividing the absolute stellar mass created in the last SFH block by
the real time spanned by the block.  We found that the
pseudo-instantaneous SFR for the high mass sources changed from
$75\pm26$\,M$_\odot$\,yr$^{-1}$ at high redshifts to
$20\pm5$\,M$_\odot$\,yr$^{-1}$ at low redshifts. In comparison, the
change for the low mass sources is from from
$43\pm23$\,M$_\odot$\,yr$^{-1}$ at high redshifts to
$9\pm2$\,M$_\odot$\,yr$^{-1}$ at low redshifts. The conclusion is
therefore that we detect no significant difference in the evolution of
the pseudo-instantaneous SFR between the high and low mass sources. To
detect an absolute trend such as this, more sources would be required
to enable finer binning in mass and redshift.

An interesting point to note is that the rate of formation of stellar
mass, which is highest at early and at late times in
the high mass, high redshift sub-sample, is very similar to that
measured by \citet{dye_et_al08} for 850$\,\mu$m selected sources.
This is perhaps not too surprising given the large overlap of this
sub-sample with the 850$\,\mu$m sample which has a median value of
redshift and $\log_{10}$(M$/$M$_\odot)$ of $1.6\pm1.0$ and
$11.5\pm0.5$ respectively, where the errors give the standard
deviation. The fact that such a large stellar population was already
in place at higher redshifts suggests that the peak star formation
rate occurred significantly earlier in the history of the Universe for
high mass systems than for low mass systems. This behaviour was
observed by \citet{heavens04} for optically selected galaxies.

\begin{figure}
\epsfxsize=83mm
{\hfill
\epsfbox{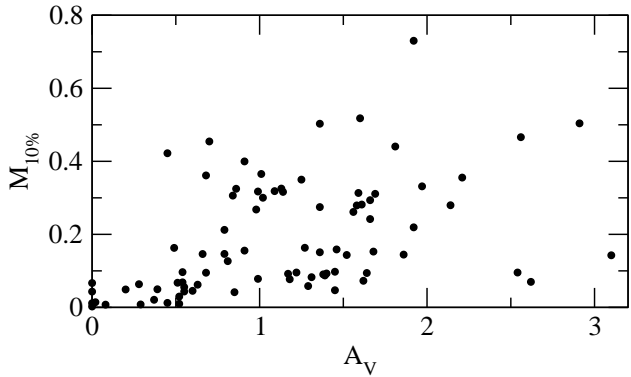}
\hfill}
\caption{The relationship between the fraction of mass formed in the
last 10\% of each galaxy's history, M$_{10\%}$, and $A_V$.}
\label{f5mass_vs_Av}
\end{figure}

To verify the robustness of our results, we conducted a series of
tests. The first was to see if the inferred SFHs are intrinsic or
merely the effect of reddening. For example, an intrinsically
late-type galaxy with strong reddening could give rise to a
reconstructed SFH with artificially suppressed late star formation.
We therefore plotted the fraction of mass formed in the last 10\% of
each galaxy's history, M$_{10\%}$, against $A_V$.  Since late activity
strongly dominates the shape of the observed SED, M$_{10\%}$ is a
sensitive indicator of SED type.  Therefore, a strong degeneracy
between the inferred lateness of an SED and extinction would manifest
itself as an obvious positive correlation between M$_{10\%}$ and $A_V$
(if the SED is more reddened, more late stellar mass is required to
maintain a fit to the observed photometry).  Figure \ref{f5mass_vs_Av}
shows these two quantities plotted for all galaxies in our sample. The
scatter is large, although there is some evidence of a correlation.
For example, all galaxies which form less than 10\% of mass in the
last 10\% of their history have values of $A_V<0.5$. However, this is
at least partly explained by the simple fact that elliptical galaxies
tend to have little or no dust. We therefore conclude that any
degeneracies between reddening and reconstructed SFHs do not
significantly affect the results described in this letter.

The second test addresses the concern that there may be a potential
bias introduced by including the mid-IR photometry based on
redshift. We therefore isolated all sources whose SED fitting used
mid-IR photometry and repeated the fitting without it. Within the
errors, which were made larger by the lack of mid-IR photometry,
we found negligible differences in the results.

As a third test, we repeated the analysis using the stellar SED
library of \citep{bruzual03}. The only significant difference we found
was that the stellar masses were an average of 40\% higher than those
computed with the Maraston SEDs.  The reconstructed SFHs of the high
mass sources also showed slightly higher fractions of mass at early
times, but this was within the errors.

Our final test was to repeat the analysis without the four sources
flagged as having strong AGN activity. The differences were again
negligible.

\section{Summary}
\label{sec_summary}

Using optical, near-IR and mid-IR photometry, we have reconstructed
the SFHs of a sample of 92 submm sources detected by BLAST. Their
SFRs peak at late times on average, consistent with the high
instantaneous SFR inferred from their submm emission.  We divided the
sample by mass and redshift into four sub-samples. Our findings
clearly indicate that the low mass sources form a much higher fraction
of their stellar mass at late times than the high mass sources,
consistent with the notion that the SFRs of higher mass galaxies
peaked at earlier times. Furthermore, the SFHs of the higher mass
sources evolve more strongly than the SFHs of lower mass sources. This
behaviour is synonymous with downsizing observed in optically-selected
samples of galaxies. Finally, our high mass, high redshift sub-sample
shows evidence of stellar mass being formed predominantly at late and
at early times, but less so when the galaxies are middle-aged. The
same trend was also observed in the sample of 850$\,\mu$m selected
sources by \citet{dye_et_al08}, although this is perhaps not
surprising given the similar range of masses and redshifts in the
sub-sample.

This letter has analysed the optical counterparts to approximately one
third of the full BLAST detected sample of galaxies presented in
\citet{dye09}. It is now understood that a significant fraction of the
galaxies detected solely at 500$\,\mu$m, are the result of flux
boosting and therefore probably not real \citep[see][]{moncelsi10}.
Our sample of 92 galaxies contains five 500$\,\mu$m-only detections.
Based on the findings of Moncelsi et al., we expect that only $\sim 1$
of these is not real.  Of the BLAST sources believed to be real,
around 50\% of these were not identified with radio and/or 24$\,\mu$m
counterparts.  Whilst some of these may have been detected in the
optical/near-IR surveys used in this letter, it is likely that the
majority will be heavily dust obscured systems lying at higher
redshifts \citep[see][]{dye09}.

Although we have allowed for extinction by dust, there could be
regions in the BLAST galaxies completely obscured at rest-frame
optical/near-IR wavelengths.  \citet{dye_et_al08} made a correction
for this effect using far-IR/submm bolometric luminosity, finding that
fully obscured star formation could result in the generation of up to
an additional 50\% of the galaxy's total stellar mass. It is likely
that this fraction is even larger for those BLAST galaxies with
radio/24$\,\mu$m counterparts but no optical/IR counterparts.

The findings described in this letter represent a taster of what would
be possible with a significantly larger sample of sources.
Although the BLAST data currently offer the largest and most
thoroughly processed sample of galaxies selected over the wavelength
range of 200 to 600$\,\mu$m to date, the increased sensitivity and
resolution of the Herschel space observatory, which recently started
operation, will soon provide vastly increased numbers of sources.  This
will enable significantly reduced uncertainties and therefore much
improved constraints on models of galaxy evolution and formation.
Nevertheless, the BLAST data will still provide a very valuable
benchmark for the Herschel data and the various analyses that will
emerge for some time to come.

\begin{flushleft}
{\bf Acknowledgements}
\end{flushleft}

SD is supported by the UK Science and Technology Facilities
Council.


{}

\label{lastpage}

\end{document}